\documentclass{article}
\usepackage{arxiv}

\usepackage[utf8]{inputenc} 
\usepackage[T1]{fontenc}    
\usepackage{hyperref}       
\usepackage{url}            
\usepackage{booktabs}       
\usepackage{amsfonts}       
\usepackage{nicefrac}       
\usepackage{microtype}      
\usepackage{lipsum}
\usepackage{graphicx}
\graphicspath{ {./images/} }
\usepackage{theorem}
\usepackage{algorithm}
\usepackage{algpseudocode}

\theoremstyle{thmstyleone}%
\theoremstyle{thmstyletwo}%

\theoremstyle{thmstylethree}%

\newcommand{\figref}[1]%
{Figure \ref{#1}%
}

\algnewcommand\algorithmicforeach{\textbf{for each}}
\algdef{S}[FOR]{ForEach}[1]{\algorithmicforeach\ #1\ \algorithmicdo}

\author{
 Rui Portocarrero Sarmento\\
  PRODEI\\
  FEUP, University of Porto \\
  \texttt{mail@ruisarmento.com} \\
   \And
 Douglas O. Cardoso\\
  COENC-PT / PPCIC\\
  CEFET-RJ \\
  \texttt{douglas.cardoso@cefet-rj.br} \\
  \And
 Pavel Brazdil \\
 LIAAD, INESC TEC \\
  \texttt{pbrazdil@inesctec.pt} \\
  \And 
 João Gama \\
 LIAAD, INESC TEC \\
 \texttt{j.gama@inesctec.pt}
}

\title{Contextualization for the Organization of Text Documents Streams}

\begin{document}

\maketitle










\begin{abstract}
   There has been a significant effort by the research community to address the problem of providing methods to organize documentation with the help of information Retrieval methods. In this report paper, we present several experiments with some stream analysis methods to explore streams of text documents. We use only dynamic algorithms to explore, analyze, and organize the flux of text documents. This document shows a case study with developed architectures of a Text Document Stream Organization, using incremental algorithms like Incremental TextRank, and IS-TFIDF. Both these algorithms are based on the assumption that the mapping of text documents and their document-term matrix in lower-dimensional evolving networks provides faster processing when compared to batch algorithms. With this architecture, and by using FastText Embedding to retrieve similarity between documents, we compare methods with large text datasets and ground truth evaluation of clustering capacities. The datasets used were Reuters and COVID-19 emotions. The results provide a new view for the contextualization of similarity when approaching flux of documents organization tasks, based on the similarity between documents in the flux, and by using mentioned algorithms.
\end{abstract}





\section{Introduction}

Previous research focused on the lower complexity of document-term matrix processing. IS-TFIDF (Incremental Similarity Term Frequency Document Frequency) and ITR (Incremental TextRank) were developed to provide a way to cope with document streams analysis and organization. The goal was to process high-dimensional matrices of documents and terms with a mapping to 2D evolving networks of documents and terms. This mapping provided faster processing of a vast amount of documents when compared to similar batch algorithms to calculate the similarity between documents.

The similarity between documents serves as a base point to organize documents of similar content in groups, i.e., clusters of documents. Therefore, we aim to discover which of the two developed algorithms provides better clustering quality by providing faster access to related documents in searches for a particular subject of study. 

For this document, we provide a throughout analysis with two large datasets with thousands of long text documents in the form of a flux of documents.

This paper is organized as follows. The section~\ref{SOTA} provides a background for the algorithms developed in previous parts of the research. Section~\ref{CSE} provides the datasets and methodology used in the case study. Section~\ref{RES} provides the results of the performed tests to obtain clustering quality measures, and~\ref{DISC} provides a discussion for these tests and results. Finally, section~\ref{CONC} concludes this document.

\section{State-of-the-Art}\label{SOTA}


In this section we describe the building blocks of the methodology here proposed.
Although the original introductions of these tools can be found in the literature, we take the opportunity to fill some gaps regarding their properties.
Moreover, this leads to a more contextualized and self-contained detailment of our work.

\subsection{IS-TFIDF for Text Streams\label{Inc}}

\cite{DBLP:journals/snam/SarmentoCDBG21} proposed IS-TFIDF, a solution for the problem of document similarity assessment when the input documents are available not as a data set but as a data stream, what imposes harder constraints on their processing \cite{MOA}.
As a consequence, the document stream is processed iteratively in a view-once fashion:
after common text preprocessing (stopwords removal, stemming etc), the terms which compose the document are used to update the respective TF-IDF values while avoiding to reassess the entire collection of terms seen;
the same goes for document similarity, whose evaluation can be limited to pairs of documents which share a term instead of every pair of them.

In a basic setting for IF-TFIDF designated by its authors as One Document Streaming (ODS), every stream element is an unprecedented document.
Besides this, the method also works when documents can not only be created but also extended during stream processing, referred to as Several Document Streaming (SDS).
Both can be used in common real scenarios: e.g., ODS could be employed to incrementally assess the similarity between papers published in a given venue through the years; in its turn, SDS enables to continually verify the pairwise alignment of users of some social network as their collections of text posts grow \cite{cossu_review_2016}.

Internally IS-TFIDF keeps a bipartite graph:
one set of nodes regards the documents while the other represents terms;
each edge indicates that the term in one of its ends is featured in the document in the other end.
Considering that $k$ is the expected number of tokens in a document after preprocessing, and that $p$ is the number of documents seen up to a given time instant, such a graph has $p$ nodes and $pk$ edges.
Moreover, considering that $n$ is the expected number of documents which have any term in common with a given document, sparse matrices for TF-IDF and document pairwise similarities are also maintained, leading to an overall space complexity of $O(p(k + n))$.
Processing each stream element updates theses data structures, what is accomplished with a time complexity of $O(n^2q)$, given that $q$ is number of terms seen up to a given time instant.

\subsection{Incremental TextRank}

\begin{figure}[ht]
\centering
\includegraphics[width=\textwidth]{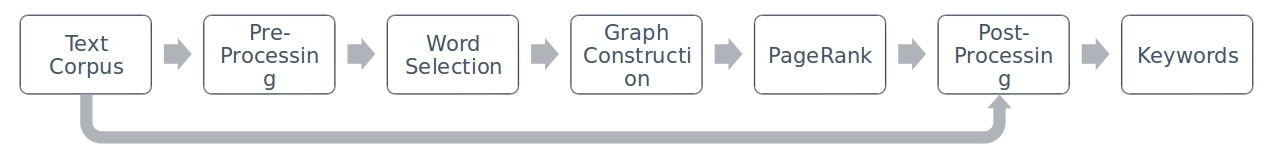}
  \caption{Original TextRank workflow (as presented in \cite{DBLP:conf/iceis/SarmentoCBG18})}
  \label{fig:TRankDiag}

\end{figure}

We describe our suggestion for two variants of an incremental TextRank by citing \cite{DBLP:conf/iceis/SarmentoCBG18} in this part. We suggest two versions: a Window-based version and a more complicated incremental version. The series of processes that the text goes through in the original TextRank is shown in \figref{fig:TRankDiag}. However, although certain components of it are rapidly expanding, such as the graph generation from text words, others are not ready for a streaming or incremental approach.

\subsubsection{Window-based Streaming TextRank}

The initial version we created was based on iterating through a data stream using a sliding window. As a result, with this version, the current snapshot processing is unconnected to the prior snapshot processing. As a result, from snapshot to snapshot, each piece of data from the stream is processed individually. We just process the current piece of data in each iteration of the window-based TextRank, we pre-process this data, and we make the normal word selection by considering only nouns and adjectives. Furthermore, as we construct the graph, we take into account the picked terms for the current piece of data. As a result, the following step of the algorithm, the PageRank method for detecting strong words in text, only examines the words in the current snapshot graph. Again, post-processing is performed with the words in the text accessible just for the present snapshot in mind. The top-$K$ phase of this window-based streaming version, as previously indicated, employs the top-$K$ space-saving approach based on the notion of landmark windows. As a result, the only information that moves from snapshot to snapshot processing for the streaming data is the top-$K$ list of keywords.

\subsubsection{Incremental TextRank}

\begin{figure}[ht]
  \centering
  \includegraphics[scale=0.53]{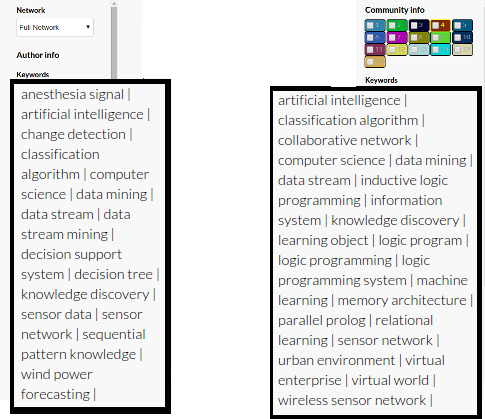}
  \caption{Keyword descriptors for a selected researcher in the left-hand side panel and for his/her affinity group in the right hand-side (extracted from \cite{trigo2015retrieval})}
  \label{fig:Keywords}
\end{figure}

The user has numerous possibilities to pick the node(s) he is looking for in the Affinity Miner Prototype, as indicated in \cite{trigo2014affinity} and \cite{DBLP:journals/snam/SarmentoCDBG21}. It may be done visually by clicking on a node or textually by searching for the researcher's name in the main tab or keywords in the secondary tab. Keywords are used to characterize each researcher's activities. The left-hand side panel displays keyword descriptions for the selected author. On the right, keyword descriptions of the researcher's affinity group are presented (Fig.~\ref{fig:Keywords}). 

The Affinity Miner prototype generates characterizing keywords of fair quality. An informal review was carried out, comparing the keywords created with the keywords collected from the researchers' web sites. For example, for the selected author, the keywords taken from his homepage (Data Mining and Decision Support; Knowledge Discovery from Data Streams; Artificial Intelligence) had a high degree of overlap with the ones derived automatically. 

\begin{figure}[ht]
\centering
\includegraphics[width=\textwidth]{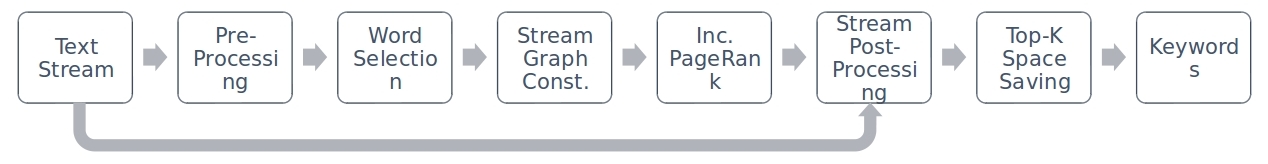}
   \caption{Incremental TextRank workflow (as presented in \cite{DBLP:conf/iceis/SarmentoCBG18})}
  \label{fig:IncTRankDiag}

\end{figure}

Nonetheless, the original TextRank algorithm's scalability was limited due to its high time complexity. As a result, we created an incremental version of this technique to reduce keyword extraction processing time and adapt it to text streams. 
\figref{fig:IncTRankDiag} depicts our concept for an improved incremental TextRank. To deal with changing text streams, we modified the original algorithm's procedure.
The input of a fresh text stream begins to be pre-processed in the same manner as the original algorithm. At this level, we filter just the required terms, such as nouns and adjectives, as we did in the original algorithm. The procedure then continues with the word graph update, progressively considering the prior graph from earlier updates, with edge weight changes for the repetition of word sequences and the insertion of new nodes/words if new words occur in the developing text stream.

The incremental PageRank method is then used to retrieve the more significant terms in the text stream. These keywords are then used to the stream's final post-processing step to obtain a suitable keyword combination, as in the original method. Finally, because we do not want to accumulate unnecessary keywords from previous stream updates, and because we want the proposed system to be sensitive to the occurrence of new keywords in the text stream, we process the keywords in a stage where we maintain and update a top-$K$ list of keywords using the previously described space-saving algorithm. 

The top-$K$ technique application, which is based on the landmark window, provides an efficient approach for large-scale text datasets. It prioritizes relevant keywords while discarding less extracted keywords over time or keywords extracted from previous snapshots. This is an alternate solution for sliding windows. The top-$K$ technique would not be applicable for \cite{Gama:2010:KDD:1855075} because it may delete important keywords from the retrieved top-$K$ keywords. Those keywords might still be included in the top-$K$ keyword list we want to keep. In our scenario, knowing the $K$ keywords of the simulated data stream from the database of publications and received after the post-processing stage of the streaming TextRank algorithms entails knowing the $K$ keywords of the simulated data stream from the database of publications and received after the post-processing stage of the streaming TextRank algorithms. The $K$ option allows the user to choose the maximum number of top-$K$ saved keywords for this application.

The {\em Space-Saving} algorithm is implemented in Algorithm \ref{topk} via the functions {\em getTopKKeywords} and {\em updateTopKeywordsList}.
New top-$K$ keywords are added to the list as the network evolves over time. Keywords that leave the top-$K$ list of keywords are deleted from the list. 
    

\begin{algorithm}
 \small
  \caption{top-$K$ Pseudo-Code for Keyword Extraction}
    \begin{algorithmic}[1]
    \renewcommand{\algorithmicrequire}{\textbf{Input:}}
    \renewcommand{\algorithmicensure}{\textbf{Output:}}
        \Require $start$, $k\_param$, $keywords$ \Comment{k parameter and previous Phase' extracted Keywords}
        \Ensure $keywords$
        
        \State $R \leftarrow \{\}$ \Comment{Current keywords}
        \State $E \leftarrow \{\}$ \Comment{Keywords currently in the Keywords' bucket}
        
        \State $R \leftarrow $ getKeywordsFromPost-ProcessingStage($start$)
        
        \While{($R <> 0$)}
        
            \ForAll {$Keywords \in R$}
                \State $before \leftarrow \Call{getTopKKeywords}{k\_param}$
                \State $\Call{updateTopKeywordsList}{keyword}$
                \State $after \leftarrow \Call{getTopKKeywords}{k\_param}$
                \State $maintained \leftarrow before \bigcap after$
                \State $removed \leftarrow before \setminus maintained$
                
                \ForAll {$keywords \in after$}  \Comment{add top-$K$ Keywords}
                    \If {$keyword \subset words$}
                        \State $\Call{addKeywordToBucket}{words}$
                        \State $E \leftarrow E \bigcup \{keyword\}$
                    \EndIf
                \EndFor
                
                \ForAll {$keyword \in removed$}  \Comment{remove non top-$K$ keywords}
                    \State $\Call{removeKeywordFromBucket}{keyword}$
                \EndFor
            \EndFor
        \EndWhile

        \State $keywords \leftarrow E$
\end{algorithmic}
\label{topk}
\end{algorithm}
 
\subsubsection{Algorithmic Analysis} 

Algorithm~\ref{alg:itrods} is a conceptual description of the procedure each document of an input stream would be subject to when ITR is employed. For ODS, each input document is unprecedented, so that the condition in~\ref{alglnitr:newnodeods} is always true.
In the several documents stream (SDS) variant, some updates are necessary regarding $PageRank$ and $top-k$ keywords. To assess the asymptotic computational complexity of the algorithm, the following variables should be considered:
PageRank algorithm has a complexity of $O(n+m)$ and top-$K$ has a complexity of $O(K \cdot \log k)$, where $K$ is the minheap window size, and $k$ is the number of keywords to retrieve from the stream. These are the dominant complexity parcels of the ITR algorithm. Nonetheless, for SDS, PageRank complexity can be simplified to $n'$ changed tokens and $m'$ changed edges, respectively, for the changed edges in $Dgt$. Therefore, the proposed algorithm has a time complexity of $O(n+m+\log k)$ for ODS and $O(n'+m'+K\cdot \log k)$ for SDS.

\begin{algorithm}[ht]
\small
   \caption{ITR Document Processing Pseudocode (ODS) \label{alg:itrods}}
   \begin{algorithmic}[1]
      \Require{$doc$, the document to be processed}
      \Require{$Dgt$, a graph per document with each documents' text tokens}
      \Require{$Dtopk$, a top-k strutcture to store each documents' top-k keywords}
      \If{there is no node regarding $doc$ in $Dgt$ \label{alglnitr:newnodeods}}
         \State Add node $token$ to $Dgt$
         \State Calculate the initial PageRank of $token$ regarding $doc$ \label{alglnitr:pr1}
         \State Calculate top-k keywords of $doc$ considering each document initial PageRank table \label{alglnitr:itopkods}
      \EndIf
      \State Preprocess $doc$
      \ForEach{token $t \in doc$ \label{algln:fortitr1}}
         \If{there is no node regarding $t$ in $Dgt$}
            \State Add node $t$ to $Dgt$
         \EndIf
         \If{there is no edge in $Dgt$ between nodes $t$ and $tn$}
            \State Add an edge between $t$ and $tn$ to $Dgt$
         \EndIf
         \State Calculate PageRank of $t$ regarding $doc$
         \label{alglnitr:pr2}
         \State Calculate list of $k$ keywords retrieved from the PageRank table for tokens of $doc$
         \label{alglnitr:topk}
      \EndFor
   \end{algorithmic}
\end{algorithm}

\begin{algorithm}[ht]
\small
   \caption{ITR Document Processing Pseudocode (SDS) \label{alg:itrsds}}
   \begin{algorithmic}[1]
      \Require{$doc$, the document to be processed}
      \Require{$Dgt$, a graph per document with each documents' text tokens}
      \Require{$Dtopk$, a top-k strutcture to store each documents' top-k keywords}
      \If{there is no node regarding $doc$ in $Dgt$ \label{alglnitr:newnodesds}}
         \State Add node $token$ to $Dgt$
         \State Calculate the initial PageRank of $token$ regarding $doc$ \label{alglnitr:prsds1}
         \State Calculate top-k keywords of $doc$ considering each document initial PageRank table \label{alglnitr:itopksds}
      \EndIf
      \State Preprocess $doc$
      \ForEach{token $t \in doc$ \label{algln:fortitr2}}
         \If{there is no node regarding $t$ in $Dgt$}
            \State Add node $t$ to $Dgt$
         \EndIf
         \If{there is no edge in $Dgt$ between nodes $t$ and $tn$}
            \State Add an edge between $t$ and $tn$ to $Dgt$
         \EndIf
         \State Update the incremental PageRank of $t$ regarding $doc$ considering only the new and changed tokens
         \State Update the list of keywords retrieved from the PageRank table for tokens of $doc$
         \label{alglnitr:ipr}
         \State Update top-k keywords of $doc$ considering each document seen changes in keywords \label{alglnitr:utopk}
      \EndFor
   \end{algorithmic}
\end{algorithm}

\subsection{Hierarchical Clustering, Word2Vec, and FastText}

By using the Stats R Package by the \cite{STATS}, we performed hierarchical clustering of the similarity matrices retrieved from previously explained algorithms. Hierarchical clustering is a type of clustering algorithm that performs clustering based on the proximity of points of the cluster, with several different methods, i.e., with different optimization measures for cluster forming. 
Hierarchical clustering is a well-known method for clustering data, as explained in, for example,  \cite{gordon1999classification}. It was updated and transformed numerous times in the past, and it is described in its various forms in publications like, for example, in \cite{Murtagh1985MultidimensionalCA,legendre2012numerical,murtagh2014ward}.


Word2Vec model is the representation for words in a vectorized distribution of values, as stated in \cite{bojanowski2016enriching}. The general characteristic of these representations is that semantically similar words have proximate representations in the target vector space. 
Word2Vec model comes from the idea of predicting neighbors of a word using the theory of neural networks. The main attribute of distributed representations was proposed initially by \cite{Rumelhart1986LearningRB}. The author proposed that the representations of semantically similar
words are close in the vector space. The vector representations of words are learned using the distributed Skip-gram or Continuous Bag-of-Words (CBOW), by \cite{41224}. The CBOW idea is to predict the word "in the middle" from the surrounding words. In contrast, in the Skipgram
model, the training objective is to learn vector representations that predict its
context in the same sentence.
Many previous works used bag-of-words in conjunction with Word2Vec to inspect the similarity between text chunks or documents. 


After keyword extraction from documents, this time with the use of previously developed IS-TFIDF or ITR, we needed to find similarities between documents to build a similarity graph. 
To do this, we could use the mentioned IS-TFIDF for Incremental Similarity and ITR to retrieve a list of keywords per document in the flux. We could try to contextualize the similarity between documents with external word embedding and the Word2Vec model.
FastText \cite{bojanowski2016enriching} was used, in a context of word representation, to reach for a compound similarity of two documents, and with both calculated with the extracted significant keywords at any particular moment in the stream of documents.

\section{Case Study and Evaluation}\label{CSE}
   
This section presents the method to analyze ODS implementations of IS-TFIDF or ITR regarding both comparisons. We introduce the reader to the data used for the tests. Finally, we describe the methodology to perform the comparison tests.

\subsection{Description of the Data}

In this case study, for the algorithms used in retrieving keywords, similarity for clustering, we selected a dataset publicly available for research purposes by \cite{Signal1M2016}, and to obtain corresponding measurements for evaluation of quality. We also selected another dataset with ground truth labels, the COVID-19 emotions dataset by \cite{BibEntry2021Dec}. Additionally, we start by using \cite{TED} for the comparison of IS-TFIDF with and without contextualization, by using Word2Vec and FastText embedding.


The first dataset we selected was from   \cite{Signal1M2016}, which has a high amount of information, including Reuters news and also several articles extracted from Blogs. The data contains article titles, content, type of publication (which might include news or Blog posts), the source institution, and also date and time of publishing. This structured data's high quality and high organization make it a good source for text mining or NLP tasks.
We selected Reuters news, corresponding to circa 2000 news articles. We could choose between the news Titles or the news content to analyze regarding the used text. We choose the news content in all our studies.
Thus, we used this particular dataset to test the ODS method in our experimental setup. Each document presented to the input in the flow of the stream of documents was considered a new document.


TED talks dataset \cite{TED} was used to test the hypothesis that IS-TFIDF would gain with the use of Word2Vec embedding regarding clustering quality. These datasets contain information about all audio-video recordings of TED Talks uploaded to the official TED.com website until September 21st, 2017. The TED main dataset contains information about all talks, including the number of views, number of comments, descriptions, speakers, and titles. The TED transcripts dataset contains the transcripts for all talks available on TED.com. We selected the complete transcriptions for analysis in the experimentation testbeds.

The COVID-19 emotions dataset by \cite{BibEntry2021Dec} is a ground truth dataset of emotional responses to COVID-19. Participants were asked to indicate their emotions and express these in written text. This resulted in the Real World Worry Dataset of 5000 text documents, 2500 short text, and 2500 long texts. We used only long texts in our studies.

\subsection{Methodology}

We started the tests with the hypothesis that IS-TFIDF would gain using Word2Vec to improve clustering quality. We tested this assumption with de TED dataset and for hierarchical clustering with and without using Word2Vec for 15 evaluation metrics and 30 data stream points.
Then, we proceeded with the comparison between IS-TFIDF and ITR, both with Word2Vec. The ODS IS-TFIDF and ICS implementation or the ITR implementation were then evaluated in an incremental setup for both the developed versions, with the Word2Vec, i.e., FastText embedding, being used in the majority of the time. The only time the original similarity between documents with IS-TFIDF was used - i.e., by using cosine/euclidean distance similarity - was when there was no possible calculation using Word2Vec. This can occur when words are not present in the available Word2Vec dictionaries. Therefore, we used a hybrid solution for similarity update, with contextualization given by Word2Vec, to calculate the similarity between extracted keywords from both algorithms. 

To avoid disparities between algorithms, eight keywords with higher values of TF-IDF were selected per document, in the case of IS-TFIDF. Similarly, for the ITR algorithm, eight keywords were selected from the top-k table of keywords, in the case of this second algorithm for testing and comparison purposes and per document. With ITR, in the case Word2Vec model with FastText did not return a similarity value; this was considered 0. 

Both algorithms were tested with 30 stream points within a distance of 25 documents. This is equal to saying that this represents the passing of more than 1000 documents in the stream sample for similarity (continuously updated for each new document) retrieval and corresponding clustering tasks.
We had the following number of different classes per point, and for each dataset:

\begin{table}[ht]
\centering
\caption{Count of ground truth classes for each point of data stream (30 points)}
\begin{tabular}{@{}cccc@{}}
\toprule
Data Stream Point & TED & REUTERS & COVID-19 \\ \midrule
1                 & 16  & 10      & 5        \\
2                 & 26  & 15      & 5        \\
3                 & 30  & 16      & 6        \\
4                 & 42  & 18      & 6        \\
5                 & 52  & 20      & 7        \\
6                 & 53  & 24      & 7        \\
7                 & 57  & 27      & 7        \\
8                 & 61  & 31      & 8        \\
9                 & 68  & 32      & 8        \\
10                & 70  & 32      & 8        \\
11                & 71  & 35      & 8        \\
12                & 74  & 35      & 8        \\
13                & 78  & 36      & 8        \\
14                & 81  & 36      & 8        \\
15                & 83  & 37      & 8        \\
16                & 84  & 37      & 8        \\
17                & 85  & 37      & 8        \\
18                & 86  & 39      & 8        \\
19                & 89  & 42      & 8        \\
20                & 92  & 42      & 8        \\
21                & 95  & 42      & 8        \\
22                & 98  & 42      & 8        \\
23                & 99  & 42      & 8        \\
24                & 99  & 43      & 8        \\
25                & 108 & 43      & 8        \\
26                & 110 & 43      & 8        \\
27                & 115 & 43      & 8        \\
28                & 118 & 43      & 8        \\
29                & 120 & 43      & 8        \\
30                & 120 & 44      & 8        \\ \bottomrule
\end{tabular}
\end{table}


We used the complete-linkage method \cite{complete_linkage} to cluster the obtained similarity matrices and corresponding distances between documents. Although we tested some other methods, the differences were insignificant and proved to be minor when selecting a method for this type of clustering algorithm.


To calculate evaluation metrics, we used the ClusterR R package by \cite{CR}. This package returns several metrics to evaluate and validate clusters with ground truth classes. Although there are numerous published research papers and documentation that provide results using external validation metrics, there is crescent care with selecting these metrics within the research community. There is also a crescent fear within the community that these metrics offer some bias in results that might prove them not very reliable, for example, as written by \cite{10.1016/j.patcog.2016.12.003}. Thus, we choose the option to use most metrics available to make conclusions of which algorithm would be better to achieve document similarity. The list of metrics used in this study is as follows:

\begin{itemize}
    \item purity (PUR);
    \item entropy (E);
    \item normalized mutual information (NMI);
    \item variation of information (VI);
    \item normalized variation of information (NVI);
    \item specificity (SPE);
    \item sensitivity (SENS);
    \item precision (P);
    \item recall (R);
    \item F-measure (F1);
    \item accuracy OR rand-index (RI);
    \item adjusted-rand-index (ARI);
    \item jaccard-index (JI);
    \item fowlkes-mallows-index (FMI);
    \item mirkin-metric (MM)
\end{itemize}

\section{Results}\label{RES}

In this section, we summarize the results by stating average values for the 15 metrics taken with the 30 point sample of the documents stream. The higher values are underlined with bold digits for easier reading of these tables.

\subsection{IS-TFIDF and IS-TFIDF with Word2Vec - TED DATASET}

In Table~\ref{TEDTAB1}, we show the results for 15 evaluation metrics, for the comparison between having and not having Word2Vec (W2V) in IS-TFIDF algorithm, and what the implications are considering clustering evaluation metrics.

\begin{table}[ht]
\caption{Comparison between IS-TFIDF without and with Word2Vec embedding}
\centering
\begin{tabular}{@{}lcc@{}}
\toprule
Measure  & IS-TFIDF  & IS-TFIDF (W2V) \\ \midrule
PUR      & \bf{0.382}     & 0.318          \\ 
E        & \bf{0.329}     & 0.327          \\ 
NMI      & \bf{0.544}     & 0.495          \\ 
VI       & 4.649     & \bf{4.987}          \\ 
NVI      & 0.616     & \bf{0.667}          \\ 
SPE      & \bf{0.882}     & 0.857          \\ 
SENS     & 0.121     & \bf{0.125}          \\ 
P        & \bf{0.050}     & 0.030          \\ 
R        & 0.121     & \bf{0.125}          \\ 
F1       & \bf{0.062}     & 0.048          \\ 
RI       & \bf{0.857}     & 0.832          \\ 
ARI      & \bf{0.015}     & -0.006         \\ 
JI       & \bf{0.032}     & 0.025          \\ 
FMI      & \bf{0.072}     & 0.061          \\ 
MM       & \bf{36353.667} & 32856.333      \\ \bottomrule
\end{tabular}
\label{TEDTAB1}
\end{table}

Comparing versions of the IS-TFIDF algorithm, the following Table~\ref{TEDTAB} shows the Mann-Whitney-Wilcoxon (MWW) test results for the 15 metrics available in the experiments regarding the TED dataset.

\begin{table}[ht]
\centering
\caption{IS-TFIDF and IS-TFIDF with Word2Vec - TED DATASET}
\begin{tabular}{@{}lc@{}}
\toprule
Measure  & Significantly different? \\ \midrule
PUR      & \textbf{Yes}  \\ 
E        & No            \\ 
NMI      & \textbf{Yes}  \\ 
VI       & \textbf{Yes}  \\ 
NVI      & \textbf{Yes}  \\ 
SPE      & No            \\ 
SENS     & No            \\ 
P        & \textbf{Yes}  \\ 
R        & No            \\ 
F1       & \textbf{Yes}  \\ 
RI       & No            \\ 
ARI      & \textbf{Yes}  \\ 
JI       & \textbf{Yes}  \\ 
FMI      & \textbf{Yes}  \\ 
MM       & No            \\ \bottomrule
\end{tabular}
\label{TEDTAB}
\end{table}

As seen in Table~\ref{TEDTAB}, we have statistically significant equality of metrics for almost half of the evaluation metrics, providing some evidence that IS-TFIDF might not gain nor decrease quality with the use of Word2Vec. We then proceeded with a comparison using other datasets for both ITR and IS-TFIDF using contextualization and Word2Vec embedding.

\subsection{Quality Measurements Comparison - Reuters DATASET}

Comparing both algorithms, the following Table~\ref{reutersqualtab} shows the average value of the 15 metrics available for the Reuters dataset experiments:


\begin{table}[ht]
\caption{Quality Measurements Comparison - Reuters DATASET}
\centering
\begin{tabular}{@{}lcc@{}}
\toprule
Measure     & IS-TFIDF                    & ITR                              \\ \midrule
PUR         & \bf{0.652}                  & 0.478                            \\ 
E           & \bf{0.410}                  & 0.140                            \\ 
NMI         & \bf{0.434}                  & 0.183                            \\ 
VI          & \bf{3.838}                  & 3.533                            \\ 
NVI         & 0.721                       & \bf{0.896}      \\ 
SPE         & \bf{0.780}                  & 0.226                            \\ 
SENS        & 0.207                       & \bf{0.782}      \\ 
P           & 0.195                       & \bf{0.210}      \\ 
R           & 0.207                       & \bf{0.782}      \\ 
F1          & 0.196                       & \bf{0.326}      \\ 
RI          & \bf{0.663}                  & 0.338                            \\ 
ARI         & -0.012                      & \bf{0.004}      \\ 
JI          & 0.109                       & \bf{0.196}      \\ 
FMI         & 0.199                       & \bf{0.401}      \\ 
MM          & 66403.467                   & \bf{128908.000} \\ \bottomrule
\end{tabular}
\label{reutersqualtab}
\end{table}

\subsection{Quality Measurements Comparison - COVID-19 DATASET}

Comparing both algorithms, the following Table~\ref{covidqualtab} shows the average value of the 15 metrics available for the COVID-19 emotions dataset experiments:


\begin{table}[ht]
\centering
\caption{Quality Measurements Comparison - COVID-19 DATASET}
\begin{tabular}{@{}lcc@{}}
\toprule
Measure     & IS-TFIDF                    & ITR                              \\ \midrule
PUR         & 0.603                       & \bf{0.614}      \\ 
E           & \bf{0.302}                  & 0.069                            \\ 
NMI         & \bf{0.072}                  & 0.054                            \\ 
VI          & \bf{2.613}                  & 1.992                            \\ 
NVI         & 0.963                       & \bf{0.972}      \\ 
SPE         & \bf{0.269}                  & 0.060                            \\ 
SENS        & 0.714                       & \bf{0.946}      \\ 
P           & 0.390                       & \bf{0.398}      \\ 
R           & 0.714                       & \bf{0.946}      \\ 
F1          & 0.500                       & \bf{0.560}      \\ 
RI          & \bf{0.445}                  & 0.412                            \\ 
ARI         & -0.015                      & \bf{0.005}      \\ 
JI          & 0.335                       & \bf{0.389}      \\ 
FMI         & 0.525                       & \bf{0.613}      \\ 
MM          & 105590.533                  & \bf{109987.133} \\ \bottomrule
\end{tabular}
\label{covidqualtab}
\end{table}

\subsection{Comparison with Statistical Tests}

After checking the results for 15 metrics of evaluation, we concluded that the ITR algorithm provided higher metrics for the evaluation of clustering, with both datasets, in 10 of 15 possible metrics.

The list of 10 metrics are the following:

\begin{itemize}
    \item purity
    \item normalized variation of information
    \item sensitivity
    \item precision
    \item recall
    \item F-measure
    \item adjusted-rand-index
    \item jaccard-index
    \item fowlkes-mallows-index
    \item mirkin-metric
\end{itemize}

All these tests were performed using Mann-Whitney-Wilcoxon with paired samples, for the 30 sample stream points in the flux of documents, within 95\% of confidence level. Correspondingly, IS-TFIDF presented higher results of the following metrics:

\begin{itemize}
    \item entropy;
    \item normalized mutual information;
    \item variation of information;
    \item specificity;
    \item accuracy OR rand-index;
\end{itemize}

\section{Discussion}\label{DISC}

Although results are routed to ITR as a better choice for clustering with the same extracted keywords, several questions arise with this research.
One doubt is related to thresholds for the evolving networks, particularly the algorithms discarding lower values of similarity between documents. Although some ad-hoc tests were done with IS-TFIDF and similarity threshold to discard lower similarity between documents, the value was thought to be around 0.12 on a scale from 0 to 1. This is equal to say that similarity below or equal to these values was considered 0 for any particular use of the similarity value. Previous literature mentioned 0.06 to obtain better community detection from networks generated from similarity values, for example, in the affinity miner prototype. Nonetheless, we discovered that, with our test conditions, this value of 0.12 threshold provided better evaluation metrics for clustering for our datasets in these experiments. These values should be better defined to have a clear notion of differences between using community detection or clustering from similarity values.

Another doubt that is equally related to thresholds is the one that comprises the use of Word2Vec and FastText. Adhoc tests showed that a threshold of similarity between documents below 0.6 and the corresponding setting of similarity value to 0 represented better metrics for clustering quality. This also represented better values for modularity, with the tested community detection algorithms, for example, the one in \cite{Cordeiro2016}, which detects communities by optimizing modularity and is available in the DynComm R package by \cite{dyncomm}. This is equal to saying that, this time, the setting to 0.6 is equally related to better results either for modularity or general clustering evaluation metrics. Nonetheless, a better notion - for example, with a considerable amount of different datasets - how these thresholds change with different techniques for community detection or clustering analysis is yet to be discovered.

Another discussion point in this research comprises the model selected to test the hypothesis that contextualization gives better quality, even for the worst-case scenario of our algorithms. Either with IS-TFIDF or ITR, we are currently, in this part of the research, using the ODS model for the streaming flux, i.e., one document processing at any given moment of the flux of documents. This option is considered to have the lowest quality in achieving keywords from streaming documents. Thus, ITR algorithms' top-k phase behaves as a top-k selection of keywords, and Incremental PageRank is not being incremented at each new document, simply because every new document in the flux is what it is, a new one, and not an update of an already known document. Either way, IS-TFIDF, and ITR show that, even with zero historical background of any kind, for any dataset, achieve and maintain along with the flux, reasonable evaluation metrics for the clustering quality. This is related to the contextualization with Word2Vec and FastText vector dictionaries.

\section{Conclusions}\label{CONC}

Through the use of two previously developed algorithms, based on the assumption that large and high-dimensional document-term matrices could be mapped to evolving networks - bipartite, in the case of IS-TFIDF, and a network of word ranks, the incremental page rank of document words in the case of ITR - we presented the quality measurements comparison between both to provide better decision when opting for one of them. Additionally, the comparison of clustering quality is provided by using one algorithm in a hybrid solution, the IS-TFIDF. IS-TFIDF occasionally performs similarity calculations without external contextualization when Word2Vec does not present a clear value for similarity. ITR, in another way, is used with the same amount of keywords extracted from documents, with Word2Vec to estimate the similarity between documents.

ITR with Word2Vec model proved to surpass IS-TFIDF with Word2Vec model in most of the used metrics to compare clustering quality after using hierarchical clustering. Statistically, most tests were different, with either the Reuters Dataset or the COVID-19 emotions dataset serving as ground truth for the tests.

For future work, we are planning to use additional ground truth test datasets and other settings in both algorithms to study some of the hyper-parameters effects on clustering quality for these document organization systems with used streaming algorithms.

\section*{Acknowledgments}
This work was fully financed by the Faculty of Engineering of the Porto University. Rui Portocarrero Sarmento also gratefully acknowledges funding from FCT (Portuguese Foundation for Science and Technology) through a Ph.D. grant (SFRH/BD/119108/2016). The authors want to thank also to the reviewers for the constructive reviews provided in the development of this publication.

\bibliographystyle{unsrt}  
\bibliography{Report}

\end{document}